\documentclass[twocolumn]{aastex631}

\usepackage{url}

\usepackage{amsmath,bm}
\usepackage{float}
\usepackage{rotating}
\usepackage{gensymb}
\usepackage{hyperref}
\usepackage{xcolor}
\usepackage{multirow}
\graphicspath{{./}{Figures/}}

\shorttitle{Stability and Gini index}
\shortauthors{Wu et al.}

\begin{document}
\title{Enhanced Stability in Planetary Systems with Similar Masses}

\author[0000-0001-9424-3721]{Dong-Hong Wu}
\affiliation{Department of Physics, Anhui Normal University, Wuhu Anhui, 241000, PR, China}
\author[0000-0002-9063-5987]{Sheng Jin}
\affiliation{Department of Physics, Anhui Normal University, Wuhu Anhui, 241000, PR, China}
\author[0000-0003-2202-3847]{Jason H. Steffen}
\affiliation{University of Nevada, Las Vegas, Department of Physics and Astronomy, 4505 S Maryland Pkwy, Box 454002, Las Vegas, NV 89154, USA}

\correspondingauthor{Dong-Hong Wu}
\email{wudonghong@ahnu.edu.cn}

\begin{abstract}
This study employs numerical simulations to explore the relationship between the dynamical instability of planetary systems and the uniformity of  planetary  masses within the system, quantified by the Gini index. Our findings reveal a significant correlation between system stability and mass uniformity. Specifically, planetary systems with higher mass uniformity demonstrate increased stability, particularly when they are distant from first-order mean motion resonances (MMRs). In general, for non-resonant planetary systems with a constant total mass, non-equal mass systems are less stable than equal mass systems for a given spacing in units of mutual Hill radius. This instability may arise from the equipartition of the total random energy, which can lead to higher eccentricities in smaller planets, ultimately destabilizing the system. This work suggests that the observed mass uniformity within multi-planet systems detected by \textit{Kepler} may result from a combination of survival bias and ongoing dynamical evolution processes.  
\end{abstract}


\section{Introduction} 
\label{section:intro}

NASA's \textit{Kepler} \citep{Borucki2010} mission has uncovered a multitude of multi-planet systems characterized by compact and coplanar configurations \citep{Lissauer_2011,Fabrycky_2014}. Notably, these systems exhibit a remarkable tendency---the planets within them often share similar sizes and planetary masses, a phenomenon affectionately termed the `peas-in-a-pod' trend \citep{Weiss_2018,Millholland_2017,Wang_2017,Goyal_2022,Otegi2022}.  Various mechanisms have been proposed to explain the origins of the `peas-in-a-pod' trend, which generally fall into two frameworks: planet formation and dynamical instability. In the former scenarios, attention is directed towards the early stages of planet assembly \citep{Adams2019,Adams2020,Mishra2021,Batygin2023}, where the planet formation process tends to yield planets with comparable masses. Conversely, the latter framework suggests that post-formation dynamical processes naturally lead to the observed intra-system uniformity through frequent planetary collisions, even if such uniformity was absent initially  \citep{Goldberg_2022,Lammers2023,Ghosh2024}. 

In this context, we propose an alternative viewpoint: multi-planet systems with relatively uniform planetary masses are more likely to have extended stable lifetimes, increasing their detectability.  The dynamical instability of multi-planet systems has been extensively studied. A key factor that determines the stability of a particular system is the mutual separation $K$ between neighboring planets, which is defined as follows:
\begin{equation}
    K=\frac{a_{i+1}-a_i}{R_H},\label{eq:K}
\end{equation}
where $a_i$ and $a_{i+1}$ represent the semi-major axes of consecutive planets (from inner to outer). The mutual Hill radius $R_H$ is defined as:
\begin{equation}
    R_H=\frac{a_{i}+a_{i+1}}{2}\left (\frac{m_{i}+m_{i+1}}{3M_{\star}} \right )^{1/3}, \label{eq:R_H}
\end{equation}
where $m_i$ and $m_{i+1}$ are the masses of the respective planets, and $M_{\star}$ is the mass of the host star.  

Historically, most research \citep{Chambers1996,Zhou_2007,Smith2009,Funk2010,Obertas2017,Rice2018} has concentrated on systems with equal mutual separation (i.e. equal $K$) and equal mass, finding that the dynamical instability timescale $\tau$ (the physical timescale $t$ scaled by the orbital period of the inner most planet $t_0$) is proportional to $K$. 

\citet{Rice2023} investigates the dynamical instabilities in planetary systems with equal separations but non-equal masses, finding that the relationship between $\tau$ and $K$ persists as long as mass deviation remains below 30\%. However, the study did not address how varying degrees of mass non-uniformity affect dynamical instability times. In this work, we explore the correlation between the dynamical instability timescale of planetary systems and their intra-system mass uniformity.

Section \ref{sec:simulation setup} outlines the basic setup of our simulations.  Section \ref{sec:dynamical instability} discusses the relationship between the dynamical instability timescale and mass heterogeneity. A discussion on the dynamical instability of near resonant systems are given in Section \ref{sec:discussion}. Section \ref{sec:conclusion} provides a brief conclusion to our work.

\section{Simulation setups}\label{sec:simulation setup}

We simulated a series of planetary systems with equal separation in units of the Hill radius. For each system, the spacing parameter $K$ was assumed to be constant. We explored different values of $K$, ranging from 3 to 10. For $K\leq8$, $K$ was incremented by 0.05; for $8<K\leq10$, due to computational constraints, we simulated only 15 distinct $K$ values, uniformly distributed between 8 and 10.

Each planetary system consisted of eight planets with initially circular and coplanar orbits around a solar mass star, with the total mass of the planets fixed at 30 Earth masses ($m_{\oplus}$). In line with \citet{Goyal_2022}, we used the adjusted Gini index ($\mathcal{G}_m$) as an indicator of the mass heterogeneity in each system. The formula for $\mathcal{G}_m$ is:
\begin{equation}
    \mathcal{G}_m=\frac{1}{2N_{\rm pl}(N_{\rm pl}-1)\bar{m}}\sum_{i=1}^{N_{\rm pl}}\sum_{j=1}^{N_{\rm pl}}|m_i-m_j|,\label{eq:G_m}
\end{equation}
where $N_{\rm pl}$ represents the number of planets in each system, $m_i$ denotes the planetary mass of the $i_{\rm th}$ planet, and $\bar{m}$ is the average planetary mass in each system. A $\mathcal{G}_m$ value of 0 indicates identical mass across all planets, whereas a value approaching 1 suggests a dominant mass held by a single planet. We designed a variety of mass distributions, ensuring each planetary mass was larger than 0.1 $m_{\oplus}$, and achieved a maximum $\mathcal{G}_m$ of approximately 0.97. For $K \leq 8$, we generated 500 systems per $K$ value, with $\mathcal{G}_m$ values uniformly spread between 0 and 0.97. For $8 < K \leq 10$, we produced 250 systems for each $K$ value, maintaining the same distribution of $\mathcal{G}_m$.  

The subsequent step involved determining the orbital configurations of the planets.  The innermost planet was placed at a semi-major axis of 0.05 AU, and the remaining planets were positioned according to Equations \ref{eq:K} and \ref{eq:R_H}. The mean anomaly of the planets were randomly distributed between 0 and $2\pi$.

For comparison, we also simulated an additional group of planetary systems where each had equal separation and equal planetary mass. For each of the K values previously specified, we generated 100 planetary systems. Each system consisted of eight planets with equal masses, maintaining a total planetary mass of 30 $m_{\oplus}$ per system. The distribution of their orbital elements was consistent with the non-equal mass simulations.

For all planetary systems, the dynamical integration was performed using the REBOUND N-body code \citep{Rein2012} with the Mercurius integrator \citep{Rein2019}. An initial time step of $t_0/100$ was chosen, where $t_0$ denotes the orbital period of the innermost planet. Integration continued until a pair of planets experienced a close encounter or the maximum integration time of $10^8$ years was reached. A close encounter was defined as occurring when the distance between any two planets became smaller than the Hill radius of the innermost planet, expressed as $R_h=a_1(\mu_1/3)^{1/3}$, where $\mu_1$ represents the mass ratio between the innermost planet and the central star. The time of close encounter was recorded as the dynamical instability time $t$. 


\section{Dynamical instability time}\label{sec:dynamical instability}

\subsection{Equal mass systems}
\begin{figure*}[htbp]
    \centering
    \includegraphics[width=2\columnwidth]{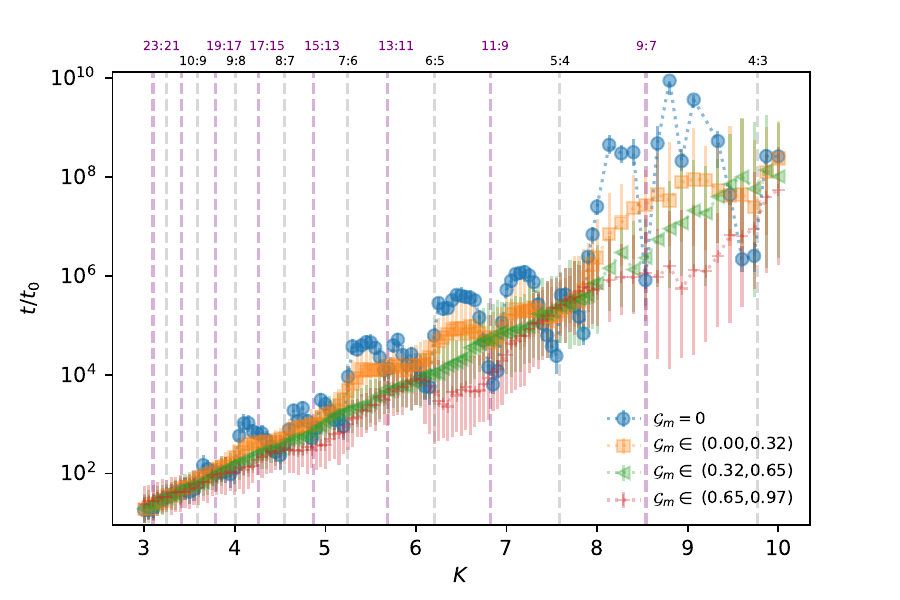}
    \caption{Dynamical instability timescale $t/t_0$ as a function of the spacing parameter $K$. The error bars are given by the 16$\rm th$ and 84 $\rm th$ percentile of the dynamical instability timescale distribution. Blue dots represent results from equal mass systems, while the orange, green and red markers correspond to systems with modified Gini index $\mathcal{G}_m$ within the ranges [0, 0.32], [0.32, 0.65] and [0.65, 0.97], respectively.  Gray and purple vertical dashed lines indicate the locations of first and second order MMRs for planet pairs in equal mass systems, respectively. } 
    \label{fig:t_K}
\end{figure*}

In Figure \ref{fig:t_K}, we present the scaled dynamical instability timescale ($t/t_0$) for both the equal mass systems and non-equal mass systems. Consistent with prior investigations \citep{Chambers1996,Obertas2017}, our results for equal mass systems support a linear relationship between $\log (t/t_0)$ and $K$. However, deviations from this trend are observed near first and second-order mean motion resonances (MMRs), marked by vertical dashed lines in Figure \ref{fig:t_K}. These deviations can be attributed to the influence of significant MMR overlap, which induces chaotic diffusion, as explored in earlier research \citep{Murray1997, Deck2013, Obertas2017}.

\subsection{Non-equal mass systems}

\subsubsection{Impact of Gini index $\mathcal{G}_m$}
For non-equal mass systems, we categorize them into three groups based on the value of $\mathcal{G}_m$. Our simulations unveil a positive correlation between dynamical instability times and the separation parameter $K$ across all groups. However, the pronounced dips in instability times near MMRs, evident in equal mass systems, diminishes in non-equal mass systems with $\mathcal{G}_m\geq 0.32$, aligning with findings from \citet{Chambers1996} and \citet{Rice2023}.  This discrepancy arises because non-equal mass systems typically exhibit unequal period ratios between neighboring planet pairs, reducing the likelihood of MMR overlap. As $K$ increases, there is a tendency that the disparity between equal mass and non-equal mass systems becomes more pronounced.


A notable trend observed is that the instability timescale decreases as the Gini index $\mathcal{G}_m$ increases, especially when the planetary configurations are distant from first-order MMRs. For systems with $K<=4$, the influence of $\mathcal{G}_m$ on the instability time is negligible due to the high density of MMRs. However, for systems with $K>4$, the impact of $\mathcal{G}_m$ becomes significant. Typically, a higher $\mathcal{G}_m$ correlates with shorter dynamical instability timescales.  

\begin{figure*}
    \centering
    \includegraphics[width=2\columnwidth]{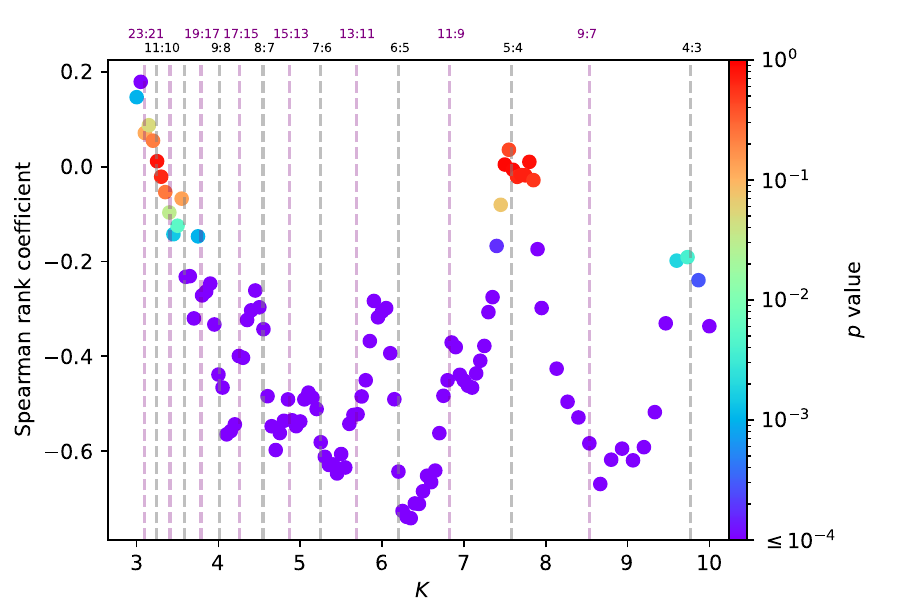}
    \caption{Correlation coefficients and $p$ values of the Spearman Rank correlation between $\log{(t/t_0)}$ and $\mathcal{G}_m$. }
    \label{fig:corr}
\end{figure*}

Analysis of eccentricity evolution indicates that smaller planets develop eccentricity faster, leading to earlier close encounters.  This could be explained by the equipartition of the total random energy ($(1/2)mv^2$) of planets \citep{Kokubo2012}. $v$ is the random velocity of planets, given by

\begin{equation}
    v\approx \sqrt{e^2+i^2}v_K,
\end{equation}
where $v_K$ is the Kepler velocity of the local circular coplanar orbit. $e$ and $i$ are the orbital eccentricity and inclination of the planets, respectively. 

In our simulations, all planets started out in coplanar orbits, thus $i$ remains to be zero throughout the simulations. The random energy could be approximated as $(1/2)me^2/a$. In other words, $e\propto m^{-1/2}$. This is similar to the concept that angular momentum dissipation (AMD) targets the smallest planets, driving their eccentricities higher and making the system less stable, which has been studied by \citet{Lithwick2014PNAS}. When the orbital eccentricity $e$ is not large, the AMD can be approximated as ${m\sqrt{a}e^2/2}$, closely resembling the form of random energy. Equipartition of the AMD also tends to excite the eccentricities of lower-mass planets more than those of higher-mass planets \citep{Wu_2011}. Therefore, close encounters are more frequent among lower-mass planets, supporting the trend where higher $\mathcal{G}_m$ values correlate with reduced stability.

In specific instances where $K$ values align with MMRs like $4:3$ and $5:4$, equal mass systems show shorter instability times than non-equal mass systems. This observation suggests that mass variations causing deviations from exact period ratios may increase system stability by moving planet pairs away from precise MMR locations.

Within the non-equal mass populations, we note that for cases where $K$ values correspond to the $4:3$ and $5:4$ MMRs in equal mass systems, the dynamical instability timescales across different groups with varying $\mathcal{G}_m$ values show consistency within 1 $\sigma$. Mass variations that shift planets away from exact resonance positions tend to increase system stability, whereas reductions in the minimum planetary mass enhance the likelihood of instability. In these instances, these two effects seem to balance each other out.


\begin{figure*}
    \centering
    \includegraphics[width=1.6\columnwidth]{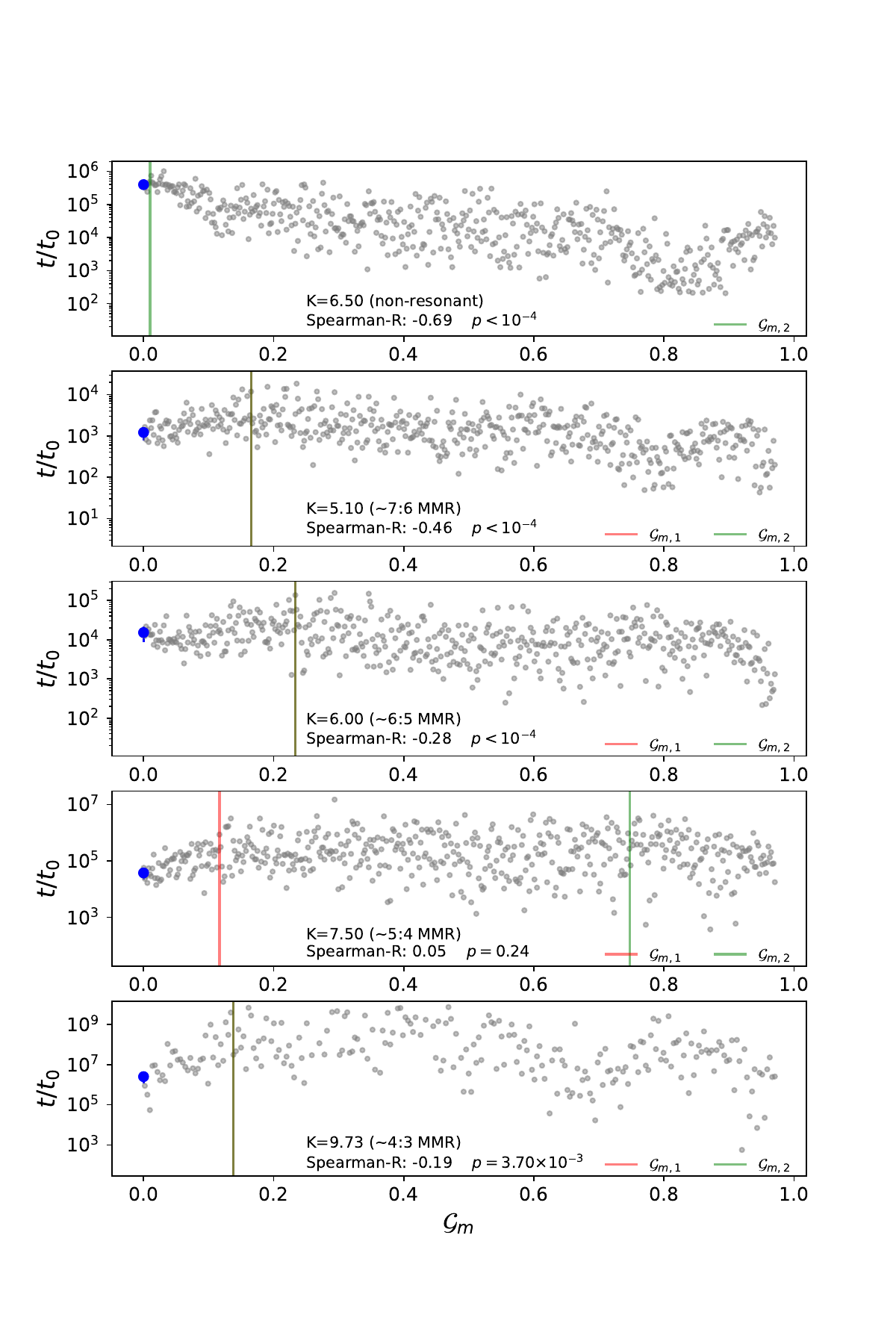}
    \caption{Dynamical instability timescale $t/t_0$ in relation to the modified Gini index $\mathcal{G}_m$ for planetary systems with $K=6.5$ (non-resonant), $K=5.1$ (near the $7:6$ MMR), $K=6.0$ (near the $6:5$ MMR), $K=7.5$ (near the $5:4$ MMR) and $K=9.73$ (near the $4:3$ MMR). The blue dots represent the median dynamical instability timescales for equal mass systems with the respective $K$ values. The gray dots represent our simulations results on non-equal mass systems. The red vertical solid line marks the critical value $\mathcal{G}_{m,1}$, where the instability timescale $t/t_0$ exhibits a positive correlation with $\mathcal{G}_m$ if $\mathcal{G}_m<\mathcal{G}_{m,1}$. The green vertical solid line represents the critical value  $\mathcal{G}_{m,2}$, where $t/t_0$ shows an anti-correlation with $\mathcal{G}_m$ for $\mathcal{G}_m>\mathcal{G}_{m,2}$. Note that the red and green vertical lines overlap in the second, third and final panel. Correlation coefficients and $p$ values for the Spearman Rank correlation between the overall $\log{(t/t_0)}$ and $\mathcal{G}_m$ are also included.}
    \label{fig:t_gini_mmr}
\end{figure*}

\subsubsection{Correlation analysis}

To analyze the correlation between the dynamical instability timescale $t/t_0$ and the Gini index $\mathcal{G}_m$, we calculate the Spearman rank coefficient between $\log(t/t_0)$ and $\mathcal{G}_m$ for planetary systems with different $K$ values, as shown in Figure \ref{fig:corr}. We find that the Spearman rank coefficients between $\log(t/t_0)$ and $\mathcal{G}_m$ in systems with $K>4$ are predominantly negative, with $p$ values $\leq 10^{-4}$,  indicating that systems with planets of similar masses tend to have longer stability periods. However, for $K$ values corresponding to the $5:4$ MMR in equal mass systems, this anti-correlation disappears. For $K$ values corresponding to the $4:3$ MMR in equal mass systems, this anti-correlation is insignificant. The correlation between $t/t_0$ and $\mathcal{G}_m$ behaves differently in planetary systems with $K$ values far away from MMRs and those near MMRs, as shown in Figure \ref{fig:t_gini_mmr}. 

To better illustrate the relationship between $t/t_0$ and $\mathcal{G}_m$, two critical values, $\mathcal{G}_{m,1}$ and $\mathcal{G}_{m,2}$, are marked in each panel of Figure \ref{fig:t_gini_mmr} as vertical solid lines in red and green, respectively. The value $\mathcal{G}_{m,1}$ is defined as the minimum value of $\mathcal{G}_m$ where the instability timescale $t/t_0$ shows a significant positive correlation with $\mathcal{G}_m$ for $\mathcal{G}_m < \mathcal{G}_{m,1}$, with $p < 10^{-4}$. Similarly, $\mathcal{G}_{m,2}$ is defined as the minimum value of $\mathcal{G}_m$ where $t/t_0$ exhibits an anti-correlation with $\mathcal{G}_m$ for $\mathcal{G}_m > \mathcal{G}_{m,2}$ (with $p < 10^{-4}$), ensuring that $\mathcal{G}_{m,2} \geq \mathcal{G}_{m,1}$ if $\mathcal{G}_{m,1}$ exists.

For planetary systems with $K=6.5$, only $\mathcal{G}_{m,2}$ exists, and we observe a notable anti-correlation between $t/t_0$ and $\mathcal{G}_m$. The Spearman rank coefficient is around $-0.69$ with $p$$<10^{-4}$, suggesting a strong inverse relationship between $\log(t/t_0)$ and $\mathcal{G}_m$. For planetary systems with $K=5.1$ (near the $7:6$ MMR), $K=6.0$ (near the $6:5$ MMR) and $K=9.73$ (near the $4:3$ MMR), the critical values $\mathcal{G}_{m,1}$ and $\mathcal{G}_{m,2}$ overlap. Dynamical instability timescale $t/t_0$ first increases and then decreases. For systems with $K$ values near the $5:4$ MMR, $t/t_0$ first increases when $\mathcal{G}_m<\mathcal{G}_{m,1}$, then remains flat when $\mathcal{G}_{m,1}\leq\mathcal{G}_{m}\leq\mathcal{G}_{m,2}$, and finally decreases when $\mathcal{G}_m\geq\mathcal{G}_{m,2}$.

When $\mathcal{G}_m\sim0$, planets in these systems have nearly equal mass, and each planet pair is very close to a first order MMR (for the near resonant cases), overlap of these MMRs tends to destabilize the system. However, if the planetary masses are modified slightly, the planet pairs shift away from the center of MMRs, making the system more stable. This explains why $t/t_0$ increases with $\mathcal{G}_m$ when $\mathcal{G}_m<\mathcal{G}_{m,1}$ in most of the systems shown in Figure \ref{fig:t_gini_mmr}. As $\mathcal{G}_m$ increases beyond $\mathcal{G}_{m,2}$, most planet pairs move away from their initial MMRs, reducing the influence of MMR overlap. In this regime, the equipartition of the total random energy tends to affect the smallest planets, increasing their eccentricities and destabilizing the system. For systems with $\mathcal{G}_{m,1}<\mathcal{G}_m<\mathcal{G}_{m,2}$, a range observed only in systems with $K$ values near the $5:4$ MMR, there is no clear correlation between $t/t_0$ and $\mathcal{G}_m$. This may result from a combination of MMR overlap effects and equipartition of random energy effects, but the exact dynamics in these systems remain unclear. As a result, the anti-correlation between $t/t_0$ and $\mathcal{G}_m$ is less pronounced for the MMR configurations compared to the non-resonant configurations.

In addition to the above simulations, we extended our numerical simulations to planetary systems with total mass of 100 $m_{\oplus}$ and 10 $m_{\oplus}$. An anti-correlation between the dynamical instability timescale and the modified Gini index $\mathcal{G}_m$ is also observed in these systems when spacing parameter $K>4$ and planets are distant from first order MMRs, aligning with previous results.

Our findings indicate that when planets are removed from MMRs, systems with planets of similar mass exhibit greater stability compared to those with significant mass disparities. The timescale for dynamical instability in systems with equal-mass planets is substantially longer—by two to four orders of magnitude—than in systems with high $\mathcal{G}_m$ values. If planets are born with large separation parameters and similar masses, such as $K>8.5$, they may survive for over 100 Myr. This enhanced stability of systems with similar-mass planets likely boosts their observability, providing an alternative explanation for the 'peas-in-a-pod' pattern observed in planetary systems.

\section{Discussion}\label{sec:discussion}

As shown in Figure \ref{fig:t_gini_mmr}, systems with $K$ values near the 6:5 and 7:6 MMRs are most stable when $\mathcal{G}_m$ values range between 0.1 and 0.3. For systems with $K$ values near the 5:4 MMR, stability is observed over a wider range, with $\mathcal{G}_m$ values between 0.1 and 0.8. In systems with $K$ values near the 4:3 MMR, the most stable systems have $\mathcal{G}_m$ values ranging from 0.2 to 0.5. However, it is demonstrated that the observed resonant planet pairs generally have $\mathcal{G}_m<0.38$, and near resonant planet pairs show enhanced mass uniformity than non resonant planet pairs \citep{Goyal2023}. This suggests a potential contradiction between our simulation results and the observation, particularly for the $5:4$ and $4:3$ MMRs. 

\begin{figure*}
    \centering
    \includegraphics[width=2\columnwidth]{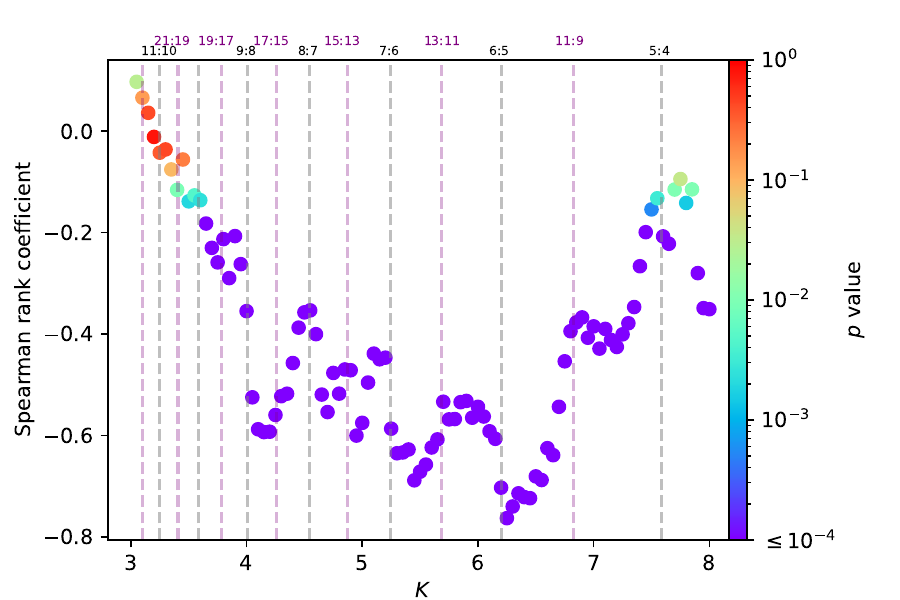}
    \caption{Correlation coefficients and $p$ values of the Spearman Rank correlation between $\log{(t/t_0)}$ and $\mathcal{G}_m$ for the well-ordered planetary systems. }
    \label{fig:corr_ordered}
\end{figure*}

It is worth noting, however, that the observed planetary systems are typically well-ordered \citep{Millholland_2017,Weiss_2018}, with outer planets tending to be larger than inner ones. In our simulations, the planetary masses are randomly distributed. To address this difference, we conducted additional simulations assuming well-ordered systems, where planetary masses increase from the inside out. To save computational time, we limited the simulations to cases with $K\leq8$, while keeping other assumptions consistent with our original simulations. We calculate the dynamical instability timescale $t/t_0$ for these new planetary systems. The Spearman rank coefficient between $\log(t/t_0)$ and $\mathcal{G}_m$ for different $K$ values of these systems is shown in Figure \ref{fig:corr_ordered}.

\begin{figure*}
\centering
    \includegraphics[width=0.8\textwidth]{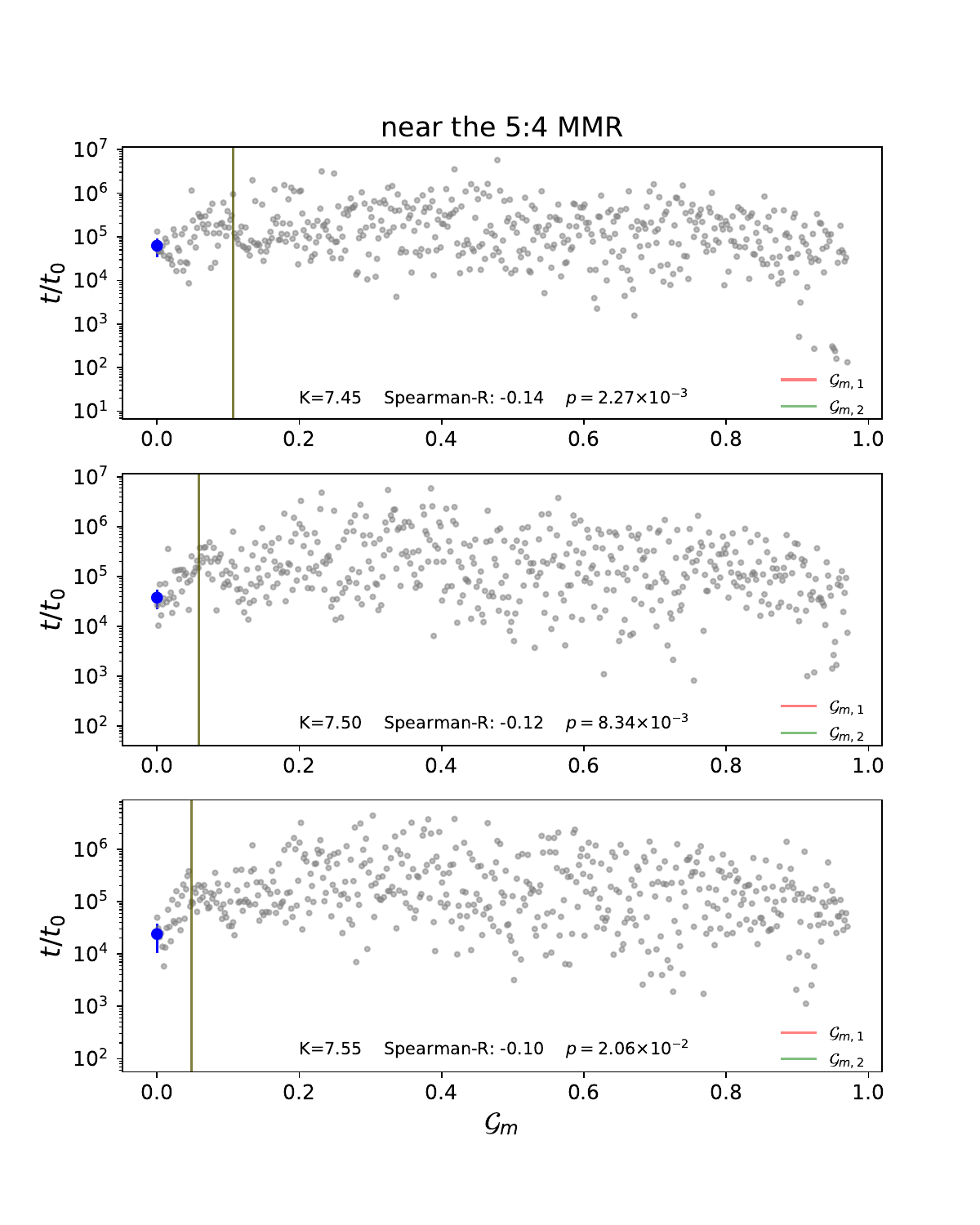}
    \caption{The relationship between the dynamical instability timescale $t/t_0$ and the Gini index of mass $\mathcal{G}_m$ for planetary systems with $K$ values near the $5:4$ MMR in the well-ordered simulations. Please refer to caption of Figure \ref{fig:t_gini_mmr} for more detail. Note that the red vertical line and green vertical line overlap in these plots. }\label{fig:54mmr_ascending}
\end{figure*}

After accounting for the ordering effect, the anti-correlation between $t/t_0$ and $\mathcal{G}_m$ remains in most cases, except for those with $K<3.8$. For planetary systems with $K$ values near the $5:4$ MMR, we observe insignificant anti-correlations between $t/t_0$ and $\mathcal{G}_m$. Figure \ref{fig:54mmr_ascending} illustrates the relationship between $t/t_0$ and $\mathcal{G}_m$ for the well-ordered planetary systems with $K=7.45$, $K=7.50$ and $K=7.55$. In these cases, $t/t_0$ first increases and then decreases, suggesting that the systems are most stable when $0.2<\mathcal{G}_m<0.4$. This range is much narrower than that derived from the original simulations. It seems that the mass ordering of planets also affects the stability of planetary systems, but we will not discuss this further in this paper. For planetary systems with $K$ values near the $6:5$ and $7:6$ MMRs, the systems are observed to be most stable when $0.1<\mathcal{G}_m<0.3$, consistent with the original simulations. As noted by \citet{Goyal2023}, near resonant planet pairs are predominantly found within the range $\mathcal{G}_m<0.38$. Our simulation results align with this observation.  

Nevertheless, we can not explain why the observed near-resonant planet pairs exhibit greater mass uniformity than non-resonant planet pairs. This discrepancy may arise from our focus on overall system uniformity rather than uniformity within individual planet pairs. Additionally, we terminate the integration of a planetary system as soon as one planet pair experiences a close encounter. However, following such encounters, other planet pairs may survive during the subsequent dynamical evolution, and collisions may occur. A more rigorous comparison with the observations would involve integrating the planetary systems over a period of more than one billion years, calculating the final $\mathcal{G}_m$ for each survived near resonant and non-resonant planet pairs, and then comparing these results with the observation. However, this approach is beyond the scope of the current work.

\section{Conclusion}\label{sec:conclusion}

In this study, we investigate the impact of mass heterogeneity on the dynamical instability time of planetary systems. Through numerical N-body simulations, we find a significant inverse correlation between the dynamical instability timescale and $\mathcal{G}_m$ when the spacing parameter $K>4$ and the planets are distant from the first order MMRs. Systems with lower $\mathcal{G}_m$, indicative of more uniform planetary masses, exhibit greater stability compared to those with higher $\mathcal{G}_m$. In general, for non-resonant planetary systems with a constant total mass, non-equal mass systems are less stable than equal mass systems when they have the same spacing parameter $K$. This instability may be attributed to the equipartition of the total random energy, which can lead to higher eccentricities in smaller planets, ultimately destabilizing the system.

Our results suggest that systems that initially possess more uniformly distributed planetary masses demonstrate enhanced stability and are thus more likely to survive longer, making them more observable. Survival bias may contribute to the observed `peas-in-a-pod' trend among \textit{Kepler} multi-planet systems. However, it can not explain the observed enhanced mass uniformity within near-resonant planet pairs as compared to non-resonant planet pairs \citep{Goyal2023}, which might be achieved through dynamical interactions and collisions \citep{Goldberg_2022, Lammers2023, Ghosh2024}.

\section{Acknowledgements}
\label{section:acknowledgements}

We thank the anonymous referee for insightful comments, which significantly enhanced this work. We also extend our gratitude to Songhu Wang for valuable discussions. This work was supported by the National Natural Science Foundation of China (NSFC; grant Nos. 12103003 and 11973094).

\bibliography{bibliography}
\bibliographystyle{aasjournal}

\end{document}